\documentclass[prd,amssymb,amsmath,amsfonts,nofootinbib,reprint,longbibliography,superscriptaddress, floatfix]{revtex4-2}

\usepackage{graphicx}
\usepackage{lmodern}
\usepackage{amsmath,amssymb}
\usepackage{mathrsfs}
\usepackage{amsfonts}
\usepackage[utf8]{inputenc}
\usepackage{url}
\usepackage[colorlinks]{hyperref}
\usepackage[dvipsnames,x11names,svgnames,rgb,table]{xcolor}
\usepackage{multirow}
\usepackage[normalem]{ulem}
\usepackage{float}
\usepackage{marvosym}
\usepackage{enumerate}
\usepackage{color,soul}
\usepackage{placeins}
\usepackage{bm}
\usepackage{color}
\usepackage{commath}
\allowdisplaybreaks
\usepackage{multirow}
\usepackage{cleveref}
\usepackage{rotating} 
\usepackage{orcidlink}
\usepackage{todonotes}
\usepackage{makecell,tabularx,colortbl}
\usepackage{booktabs,cellspace}
\usepackage{mathtools}
\usepackage{makecell}

\graphicspath{{plots/}}

\newcommand{\be}{\begin{equation}}
\newcommand{\ee}{\end{equation}}
\newcommand{\bea}{\begin{eqnarray}}
\newcommand{\eea}{\end{eqnarray}}

\newcommand{\bham}{\affiliation{School of Physics and Astronomy and Institute for Gravitational Wave Astronomy, University of Birmingham, Edgbaston, Birmingham, B15 2TT, United Kingdom}}

\newcommand{\UAM}{\affiliation{Instituto de F\'isica Te\'orica UAM/CSIC, Universidad Aut\'onoma de Madrid, Cantoblanco 28049 Madrid, Spain}}

\newcommand{\AEI}{\affiliation{Max Planck Institute for Gravitational Physics (Albert Einstein Institute), D-14467 Potsdam, Germany}}

\begin{document}

\title{Detection of GW200105 with a targeted eccentric search}

\author{Khun Sang Phukon \orcidlink{0000-0003-1561-0760}}
\email{k.s.phukon@bham.ac.uk}

\author{Patricia Schmidt \orcidlink{0000-0003-1542-1791}}
\email{P.Schmidt@bham.ac.uk} \bham

\author{Gonzalo Morras \orcidlink{0000-0002-9977-8546}}
\email{gonzalo.morras@aei.mpg.de} \AEI \UAM 

\author{Geraint Pratten \orcidlink{0000-0003-4984-0775}}
\email{g.pratten@bham.ac.uk} \bham

\begin{abstract}
  The neutron star -- black hole (NSBH) binary GW200105 was recently found to have significant residual orbital eccentricity at a gravitational-wave frequency of 20 Hz~\cite{Morras:2025xfu}. The event was originally identified  with moderate significance by matched-filter searches that employ non-eccentric templates. The neglect of relevant physical effects, such as orbital eccentricity, can severely reduce the sensitivity of the search and, consequently, also the significance of an event candidate. Here, we present a targeted eccentric search for GW200105. The eccentric search identifies GW200105 as the most significant event with a signal-to-noise ratio of $13.4$ and a false alarm rate of less than 1 in 1000 years. The best-matching template parameters are consistent with the Bayesian inference result, supporting the interpretation of GW200105 as an NSBH that formed through dynamical mechanisms including hierarchical triples and not via isolated binary evolution.
\end{abstract}

\maketitle

\section{Introduction}
\label{intro}
After the first part of the fourth observing run (O4a) by the LIGO-Virgo-KAGRA detector network~\cite{TheLIGOScientific:2014jea, TheVirgo:2014hva, KAGRA:2020tym,Aso:2013eba,Somiya:2011np}, the number of confident gravitational-wave (GW) detections has grown to more than 200~\cite{LIGOScientific:2025slb,KAGRA:2021vkt,Nitz:2021zwj,Olsen:2022pin,Mehta:2023zlk}. 
The number of detections is expected to approximately double by the end of O4~\cite{Capote:2024rmo,LIGO:2024kkz}. 
All GW signals observed to date are consistent with originating from the coalescence of compact binaries consisting of neutron star and black holes. 
The vast majority of these events were identified using templated search methods based on matched-filtering~\cite{Allen:2005fk,Cannon:2011vi,Adams:2015ulm,Luan:2011qx}. 
While this detection technique is highly efficient, many searches employed today do not yet fully incorporate important physical effects, such as spin-induced orbital precession~\cite{Lense:1918zz} and orbital eccentricity~\cite{Peters:1963ux}. 
This can lead to reduced search sensitivity and the potential loss of interesting astrophysical signals. 
Therefore, developments to incorporate these effects \cite{Nitz:2019spj, Nitz:2021vqh, Nitz:2021mzz, Dhurkunde:2023qoe, McIsaac:2023ijd, Schmidt:2024jbp, Pal:2023dyg, Phukon:2024amh, Wang:2025yac} are important and an ongoing effort in the field. 
Whilst untemplated methods could be considered~\cite{LIGOScientific:2019dag,LIGOScientific:2023lpe}, these will be less sensitive in the mass range spanned by the neutron star-black hole binaries of interest here.   
There have been several previous searches for eccentric compact binaries, though no statistically significant candidates have yet been reported. 
A targeted search for binary neutron stars was performed in~\cite{Nitz:2019spj}, though this covered a narrow mass range and did not incorporate spin effects. 
This was extended to the first search for eccentric neutron star-black hole binaries in~\cite{Dhurkunde:2023qoe} and to eccentric subsolar mass compact binaries in~\cite{Nitz:2021vqh,Nitz:2021mzz}.  
Searches for binary black hole mergers have included both templated~\cite{Pal:2023dyg,Wang:2025yac} and untemplated~\cite{LIGOScientific:2019dag,LIGOScientific:2023lpe} methods. 

Precession and eccentricity are of particular astrophysical interest as they are considered key tracers of dynamical astrophysical processes and formation pathways of the binary~\cite{Kalogera:1999tq,Samsing:2013kua,Rodriguez:2016vmx,Cholis:2016kqi,Stegmann:2025clo}. 
Elucidating the formation mechanism of the compact binaries is a major science objective in GW astrophysics~\cite{KAGRA:2021duu}. 
Recently, several independent Bayesian analyses~\cite{Morras:2025xfu,Planas:2025plq,Jan:2025fps,Kacanja:2025kpr,Tiwari:2025fua} identified residual orbital eccentricity of $\sim 0.1$ at a GW frequency of $20$Hz in the neutron star - black hole event GW200105\_162426, hereafter GW200105,~\cite{LIGOScientific:2021qlt} from the third observing run (O3). 
The event was identified by several matched-filter searches (see Table 1 in~\cite{LIGOScientific:2021qlt}) using quasi-circular template waveforms. 
Given the significant amount of residual eccentricity, this event presents an ideal case to assess the efficacy of an eccentric search on real data. 
In previous work~\cite{Phukon:2024amh}, we presented an efficient algorithm to construct an eccentric, aligned-spin template bank for low-mass binaries. 
Here, we construct a targeted search using this formalism to reanalyse publicly available strain data around the time of GW200105, and compare its performance against the default quasi-circular search. 
The paper is organized as follows: We first describe the key methodology including the data, search and significance calculation in Sec.~\ref{sec:methods}. The main results of this work are discussed in Sec.~\ref{sec:results}, where we contrast the eccentric and the quasi-circular search and demonstrate a significant improvement in the sensitive spacetime volume. We conclude in Sec.~\ref{sec:discussion}.

\section{Methods}
\label{sec:methods}
\subsection{Data}
\label{sec:data}
We analyze public O3 data~\cite{KAGRA:2023pio} from both single and coincident observation times of the LIGO Hanford-LIGO Livingston-Virgo network, comprising $\sim 8.723$  days of strain data that include the occurrence time of the event GW200105~\cite{LIGOScientific:2021qlt, KAGRA:2021vkt}. The analyzed data span from GPS time 1262192836 to 1262946499,  corresponding to UTC times from 2020-01-04 17:06:58 to 2020-01-13 10:28:01. Our analysis also utilizes data quality information~\cite{LIGO:2021ppb, Virgo:2022fxr} to avoid analyzing corrupted data and to reduce the number of false alarms due to short excess noise. No hardware injections, which are simulated detector responses to astrophysical sources, introduced by physically displacing the interferometer test masses~\cite{Biwer:2016oyg}, are present in the data~\cite{KAGRA:2023pio}. These, therefore, do not need to be accounted for in our analysis.

\subsection{Template bank and matched filtering}
\label{sec:bank}
We use the \texttt{PyCBC} pipeline~\cite{Usman:2015kfa, Nitz:2017svb, Davies:2020tsx} for matched filtering the detector data against template banks to search for GWs from compact binaries. 
We employ an aligned-spin eccentric template bank, generated using the method developed in~\cite{Phukon:2024amh}, covering the five-dimensional search parameter space consisting of the component masses, spin magnitudes and eccentricity, contrary to a four-dimensional quasi-circular parameter space used in LVK analyses~\cite{KAGRA:2021vkt, LIGOScientific:2025yae}. 
The resulting eccentric template bank contains $1,553,811$ template waveforms modeled with the frequency-domain waveform-approximant {\tt TaylorF2Ecc}~\cite{Moore:2016qxz}. The template bank is constructed using the power spectral density (PSD) of the LIGO Livingston data released from the LVK analysis~\cite{LIGOScientific:2021qlt} along with the posteriors of the eccentric analysis of GW200105 from~\cite{Morras:2025xfu}, and uses a minimal match between neighboring templates of $0.97$ evaluated in the frequency range $20-1000$ Hz. 

\begin{figure*}[t!]
 \centering
\includegraphics[width=0.49\textwidth]{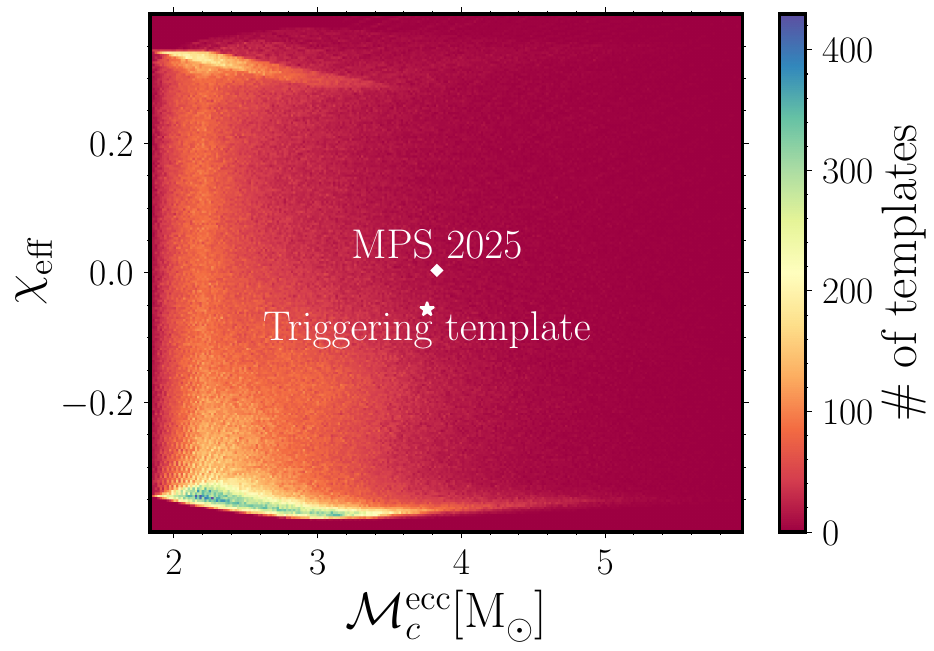}
\includegraphics[width=0.49\textwidth]{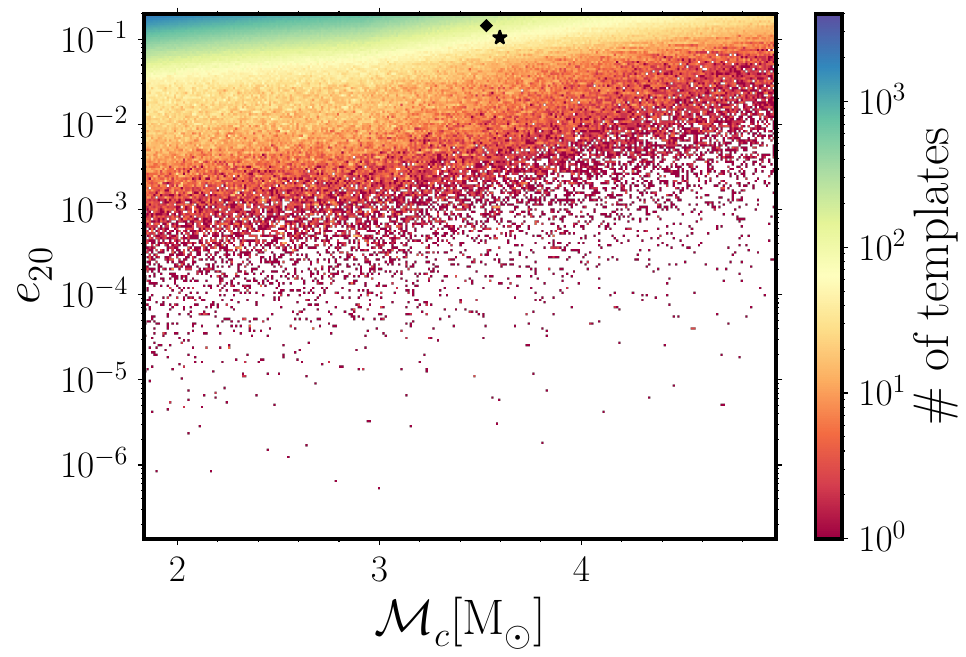}
\caption{\label{fig:template_bank} Density distribution of the templates in the eccentric template bank for the targeted search, shown in the $\mathcal{M}^{ecc}_c - \chi_{\rm eff}^{}$ and $\mathcal{M}_c^{}- e^{}_{20}$ spaces in the left and right panels, respectively. The filled diamond denotes the median posterior value for GW200105 from~\cite{Morras:2025xfu} (labeled as MPS 2025),  while the  star marks the triggering template for GW200105 in each panel.  The quasi-circular template bank spans the same parameter space as the eccentric bank in the zero-eccentricity limit.}
\end{figure*}

The span of our search parameter space is summarized in Table~\ref{tab:bank_params}. The mass and spins ranges are chosen such that they fully cover the posterior samples obtained in~\cite{Morras:2025xfu} from the reanalysis of GW200105 with an eccentric, precessing waveform model~\cite{Morras:2025nlp}. In the left panel of Fig.~\ref{fig:template_bank}, we show the distribution of templates in terms of the two dominant parameters of spin-aligned, eccentric waveforms, the effective spin, $\chi^{}_{\rm eff}$~\cite{Ajith:2009bn}, and the eccentric chirp mass, $\mathcal{M}_c^{\rm ecc}$~\cite{Favata:2021vhw}, defined as
\begin{equation}
    \mathcal{M}_c^{\rm ecc}=\frac{\mathcal{M}^{}_c}{\Big(1 - \frac{157}{24}e_{20}^2\Big)^{3/5}},
\end{equation}
where $\mathcal{M}_c = (m_1^{} m_2^{})^{3/5}/(m_1^{} + m_2^{})^{1/5}$ is the chirp mass and $e^{}_{20}$ is eccentricity at a reference GW frequency of $20$ Hz. The colors represent the density of templates, which shows the usual trend of increased density towards the lower mass region compared to higher masses. The regions of abrupt density changes correspond to the boundaries of the template placement. The right panel presents the template distribution in the $\mathcal{M}^{}_c\, – \, e^{}_{20}$ plane, showing an increase in the number of templates in regions of higher eccentricity and lower chirp mass. The median detector-frame values of GW200105 from the eccentric analysis of Ref.~\cite{Morras:2025xfu} are indicated by the filled diamonds, while the filled stars mark the best-matched template for the event in both panels.

\begin{table}[t!]
    \centering
    \begin{ruledtabular}
    \renewcommand{\arraystretch}{1.3}
    \begin{tabular}{l l}
        Parameter & Range \\
        \hline 
        Detector-frame primary mass $(m^{}_1)$ &  $[5, 15] M_\odot$ \\
        Detector-frame secondary mass $(m_2^{})$  & $[1,2.5] M_\odot$ \\
        Dimensionless primary spin magnitude $(\chi_1^{})$  &  $[0.0, 0.4]$\\
        Dimensionless secondary spin magnitude $(\chi_2^{})$  &  $[0.0, 0.05]$ \\
        Orbital eccentricity  $\left(e_{20}^{} \equiv e(20{\rm Hz}) \right)$ & $[10^{-5}, 0.2]$\\
    \end{tabular}
    \end{ruledtabular}
    \caption{Parameter space coverage of the five-dimensional eccentric template bank, where the orbital eccentricity $e_{20}^{}$ is the temporal post-Newtonian eccentricity parameter defined at a GW frequency of 20 Hz.}
    \label{tab:bank_params}
\end{table}

For a direct comparison, we conduct another matched-filter search with a quasi-circular template bank, constructing with the {\tt TaylorF2} waveform~\cite{Arun:2008kb, Buonanno:2009zt} and covering the parameter space of Table~\ref{tab:bank_params}, but with zero eccentricity. The quasi-circular template bank contains $173,581$ templates, and is generated using the geometric method of~\cite{Brown:2012qf, Harry:2013tca} with a setup otherwise identical to that of the eccentric bank. We note that the introduction of eccentricity into the search increases the bank size by roughly an order of magnitude.

As is commonly done in PyCBC~\cite{Babak:2012zx,Usman:2015kfa}, we filter the data of each detector independently against each of the two template banks to produce the signal-to-noise ratio (SNR), $\rho$, of a potential GW candidate or trigger~\cite{Allen:2005fk}. The data is sampled at a rate of 2048 Hz between 20 Hz and the Nyquist frequency. The SNR of a signal attains its optimal value when the detector noise is Gaussian and stationary, and the signal closely matches the template waveform. To reduce the impact of strong non-Gaussian features of data, we apply a correction to the SNR that encapsulates contributions from the reduced chi-square discriminator, $\chi_r^2$, that assesses the consistency of the signal with the morphology of a template waveform in the frequency domain~\cite{Allen:2004gu}. Additionally, we mitigate effects of local variations of the noise due to  non-stationarities by incorporating the variance of the PSD $v_s(t)$ at given time $t$~\cite{2020CQGra..37u5014M, Nitz:2020oeq}. These contributions rescale the SNR to produce the  reweighted SNR, $\hat{\rho}$,  
 defined as~\cite{Babak:2012zx,2020CQGra..37u5014M}
\be
\label{eq:reweighted_snr}
\hat{\rho} =  \left[ \frac{1}{2} \left\{ 1 + \left( \frac{\chi_r^2}{v_s}\right)^3\right\} \right]^{-1/6} v_s^{-1/2} \rho,
\ee
where $\chi_r^2/v_s$ and $v^{}_s$ are set to unity when $\chi_r^2/v_s\leq 1$ and $v_s\leq 0.65$, respectively~\cite{2020CQGra..37u5014M}.
Additionally, we discard triggers if either of $\chi_r^2$ and $v_s^{}$ exceeds 10 by setting $\hat{\rho}$ to unity. We use the same definition of $\rho$ and $\chi_r^2$ in the eccentric search as in the quasi-circular search, since the {\tt TaylorF2Ecc} waveform model includes only the dominant GW mode at twice the orbital frequency, similar to {\tt TaylorF2}. Concretely, he {\tt TaylorF2Ecc} waveform model does not include higher eccentric harmonics or eccentric amplitude corrections, with eccentricity only entering through the phase evolution. Due to the qualitative similarity of {\tt TaylorF2Ecc} with {\tt TaylorF2}, the traditional matched-filter SNR $\rho$ and the associated $\chi_r^2$ statistics originally developed for spin-aligned, quasi-circular waveforms accounting only for quadrupolar radiation, remain applicable for the {\tt TaylorF2Ecc} model used in this work. The traditional $\chi_r^2$-test, that essentially tracks excess power in frequency bins using the same  $\rho$ statistics might fail if signal or template contains multiple eccentric harmonics.

The reweighted SNR $\hat{\rho}$ can be further down-weighted by the Sine-Gaussian discriminator, $\chi_{r, \mathrm{sg}}^{2}$~\cite{Nitz:2017lco}, which mitigates the impact of short-duration noise transients. However, this test is  primarily relevant for searches targeting binaries with total masses greater than 40$M_{\odot}$~\cite{Nitz:2020oeq}, and therefore it does not contribute to $\hat{\rho}$ in our analysis. For a trigger from an astrophysical signal in stationary, Gaussian data that matches with a template waveform, we expect $\hat{\rho} \approx \rho$. In the quasi-circular search, this optimality might not be reached for an eccentric signal even under ideal data conditions as intrinsic differences between the signal and template morphology can yield $\chi_r^2 > 1$, leading to $\hat{\rho} < \rho$.

\subsection{Ranking statistics and significance}
\label{subsec:rankstat}
For the detection of astrophysical signals, triggers are ranked using the Neyman-Pearson statistics which reflects the relative probability of signal versus noise origin. The ranking statistics $\Lambda_s^{}$ of a trigger is, in general, given in terms of the ratio of the expected rate density of signal triggers, $R^{}_S$, to that of noise triggers, $R^{}_N$, as follows~\cite{Biswas:2012tv}:
\be
\Lambda^{}_s (\vec{k}) \propto \log R^{}_S(\vec{k}) - \log R^{}_N(\vec{k}). 
\ee
The ranking statistics depends on a set of parameters $\vec{k}$ that includes the reweighted SNRs from each detector and the optimal SNRs computed at a reference distance with the parameters of the triggering template. The signal trigger rate density $R_S^{}$ for a given template is proportional to the instantaneous sensitive or horizon volume of the detector network if sources are homogeneously distributed. The optimal SNR of a template, which is the SNR of a signal located directly overhead and perfectly matching the template, provides a measure of the template's intrinsic sensitivity. To approximate the detector network's sensitive volume for the $i$-th template, the optimal SNR from the least sensitive detector at the trigger time, $\sigma^{}_{{\rm min},i}$, is used. The median of optimal SNRs, $\bar{\sigma}^{}_{i, \rm{HL}}$ of the template in a reference detector network consisting of the LIGO Hanford (H) and LIGO Livingston (L) detectors is used to normalize the sensitive volume factor, resulting in $\left( \sigma^{}_{{\rm min},i}/\bar{\sigma}^{}_{i, \rm{HL}}\right)^3$~\cite{Nitz:2020oeq,Davies:2022thw}. In the case of a single detector observation, the $\sigma^{}_{{\rm min},i}$ term in the sensitive volume factor is replaced by optimal SNR $\sigma_i^{}$ in the detector.
In the multi-detector definition of $R_S^{}$, $\vec{k}$ also includes time delays, relative amplitudes and phases across the detectors to determine the probability of the trigger's coherence between detectors~\cite{Nitz:2017svb,Nitz:2020oeq,Davies:2020tsx}, which is absent in $\Lambda^{}_s$ for single detector observations. In practice, $\vec{k}$ also includes the template parameters $\vec{\theta}$, to further optimize $R_S$ by incorporating the density of detectable signals from astrophysical populations at a given $\vec{\theta}$~\cite{Dent:2013cva,Kumar:2024bfe}. In the analysis presented here, we do not use such an optimization and treat each template equally in detecting signals.

The noise trigger rate $R^{}_N$ is determined hierarchically for each template and detector using a model fitted to noise triggers with a decaying exponential function of $\hat{\rho}$~\cite{Nitz:2017svb}. The noise triggers used for the fitting procedure for a given template and detector are constructed by including all triggers generated by the template in the detector's data. To avoid contamination from astrophysical signals, triggers within a $0.1s$ window of the highest $\hat{\rho}$ are excluded from the noise trigger samples~\cite{Nitz:2017svb,Davies:2022thw}. At first, the noise trigger rate, $R^{}_{d,i}$, for each detector $d$ and template $i$ is determined by fitting
\be
R^{}_{d,i} (\hat{\rho})= R^{0}_{d,i} \exp [ -\alpha^{}_{d,i} (\hat{\rho}-\hat{\rho}^{}_{\rm th})],
\ee
where $\alpha^{}_{d,i}$ is the fit parameter and $R^{0}_{d,i}= N^{}_{d,i} \alpha^{}_{d,i}$ is the amplitude of the function with $N^{}_{d,i}$ triggers with $\hat{\rho}>\hat{\rho}^{}_{\rm th}$. Then, the fit parameters are smoothed over similar templates to reduce the statistical uncertainty~\cite{Nitz:2020oeq}. For the quasi-circular search, we smooth  the fit parameters of $R_{d,i}^{}$ over templates grouped in the   $\mathcal{M}^{}_{\rm c}-\chi_{\rm eff}^{}-\eta$ space, whereas the $\mathcal{M}^{\rm ecc}_{\rm c}-\chi_{\rm eff}^{}-\eta$ space is used for smoothing the  fit parameters of noise models in the eccentric search.

The noise trigger rates $R^{}_{d,i}$ from all detectors are then combined to build the model of coincidence noise rates for various detector combinations using the relation $R^{}_N = A^{}_{N\{d\}}\sum^{}_d R^{}_{d,i}$, where $A^{}_{N\{d\}}$ is time window for noise coincidence~\cite{Nitz:2020oeq,Davies:2020tsx}. In the case of a single detector trigger, the noise rate $R^{}_N$ reduces to $R^{}_{d,i}$~\cite{Davies:2022thw}. We use normalized noise rates when estimating $\Lambda_s^{}$. We note that the PyCBC ranking statistics employed in~\cite{LIGOScientific:2025yae} uses statistical data-quality information from auxiliary channels of detectors
 ~\cite{Essick:2020qpo} to update noise trigger rate, which is not incorporated here. 

For candidate events obtained in searches, generally, false alarm rates (FARs) are assigned as the measure of statistical significance. For candidates found in coincident observations, FARs are estimated by comparing their ranking statistics with the rate of ranking statistics of coincident background events constructed from each detector's triggers using the time-slide method~\cite{Babak:2012zx,Usman:2015kfa}.  
For single-detector observations, the FAR is obtained by fitting the probability density of the ranking statistics of noise triggers to a decaying exponential function and extrapolating the rate with a maximum possible inverse FAR (IFAR) of 1000 years~\cite{Davies:2022thw,LIGOScientific:2025yae}.

\section{Results}
\label{sec:results}

\subsection{Search robustness and efficiency}
\label{sec:vt}

To validate the performance of a search and assess its sensitivity to compact binary mergers, simulated signals from such sources are injected into the data and subsequently analyzed. A robust analysis should recover sufficiently loud injections that are within the search parameter space and minimally affected by data quality issues. The search sensitivity is quantified in terms of the product of the sensitive volume and time, $\langle VT \rangle$, which is proportional to the fraction of injections recovered by the search with a detection statistic above a threshold, typically given in terms of IFAR~\cite{Tiwari:2017ndi}.

We generate two populations of compact binaries with nonprecessing spins, one that is eccentric and one that is quasi-circular, each containing $\sim 30,000$ binaries covering the analysis times of the searches conducted here. The ranges of intrinsic parameters for these populations are considered such that 90\% credible intervals of mass, spin, and eccentricity posteriors reported in~\cite{Morras:2025xfu} are covered. 
Both sets of injections are uniformly distributed in distances between 40 and 1000 Mpc and isotropically on the sky. We will denote a uniform distribution in the range $(x,y)$ by $\mathrm{U}(x,y)$. The component masses in the detector frame are drawn from $m^{}_1 \in \mathrm{U}(9.45,\,12.27) M^{}_\odot$, $m^{}_2 \in \mathrm{U}(1.45,\,1.87)M^{}_\odot $ with primary spin magnitude $\chi_1^{} \in  \mathrm{U}(0,\,0.4)$, secondary spin magnitude $\chi_2^{}\in \mathrm{U}(0,\,0.05)$. The cosine of the inclination angle of a source is drawn from $\mathrm{U}(0, 1)$, while the polarization angle has a distribution $\mathrm{U}(0, \pi)$. For the eccentric population, eccentricity at 20 Hz  are distributed uniformly with $e_{20}^{} \in \mathrm{U}(0.047,0.153)$. Eccentric injections are generated using the {\tt TaylorF2Ecc} waveform approximant while quasi-circular injections use {\tt TaylorF2}.

\begin{figure*}[p]

    \centering
    \includegraphics[scale=0.7]{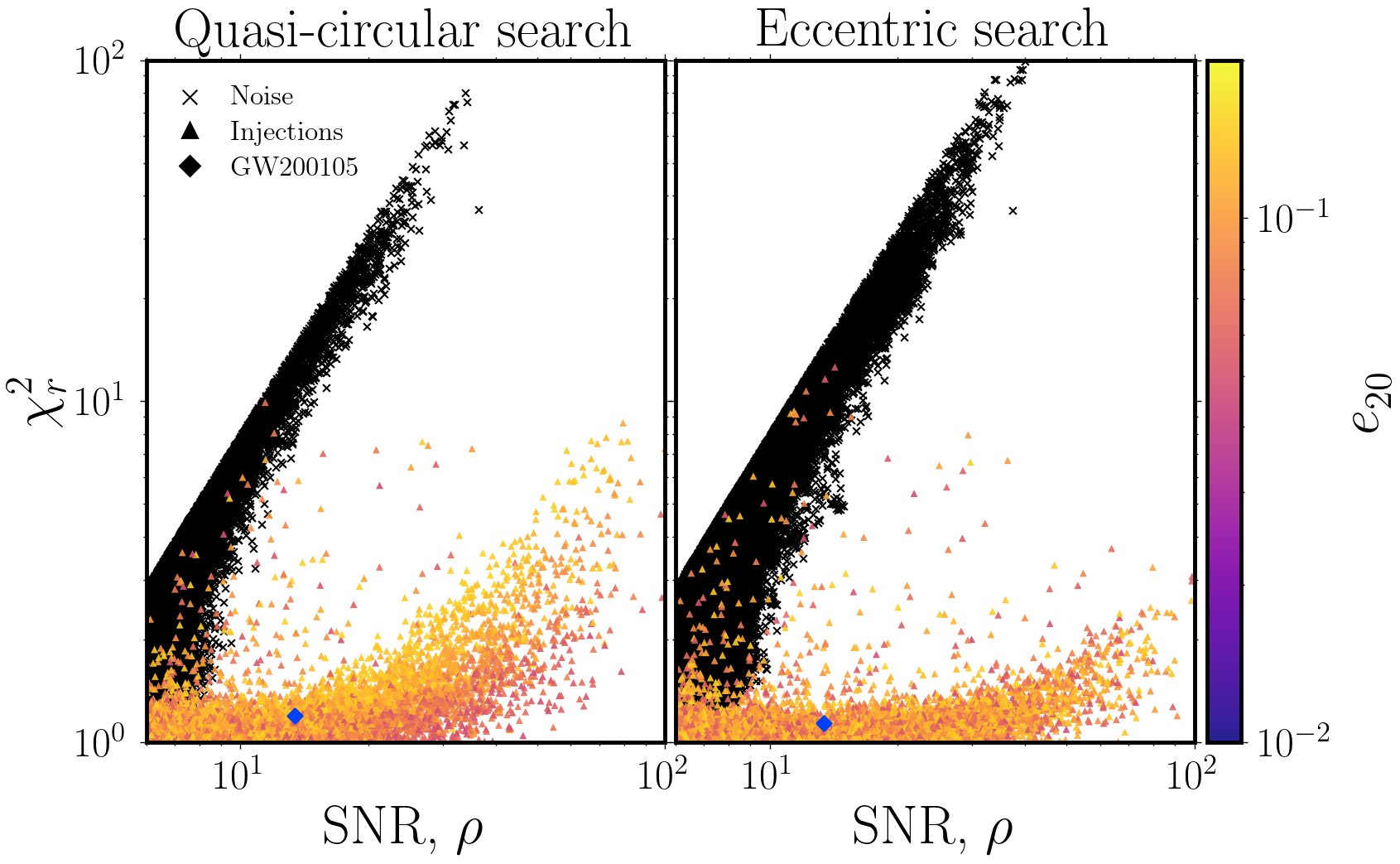} 
    \caption{\label{fig:snr_chisq} The SNR, $\rho$ and $\chi_r^2$ values of noise triggers from each detector during periods of coincident operation of at least two detectors (black), eccentric injections (colored) and the triggers for GW200105 from the quasi-circular and eccentric searches in the left and right panel, respectively. 
    The injections are colored by to the eccentricity of the injected signals at 20 Hz. The blue filled diamond markers indicate the $(\rho, \chi_r^{2})$-values of the GW200105 trigger in the eccentric and quasi-circular search, respectively. 
    }
\vspace{1cm}
    \includegraphics[scale=0.7]{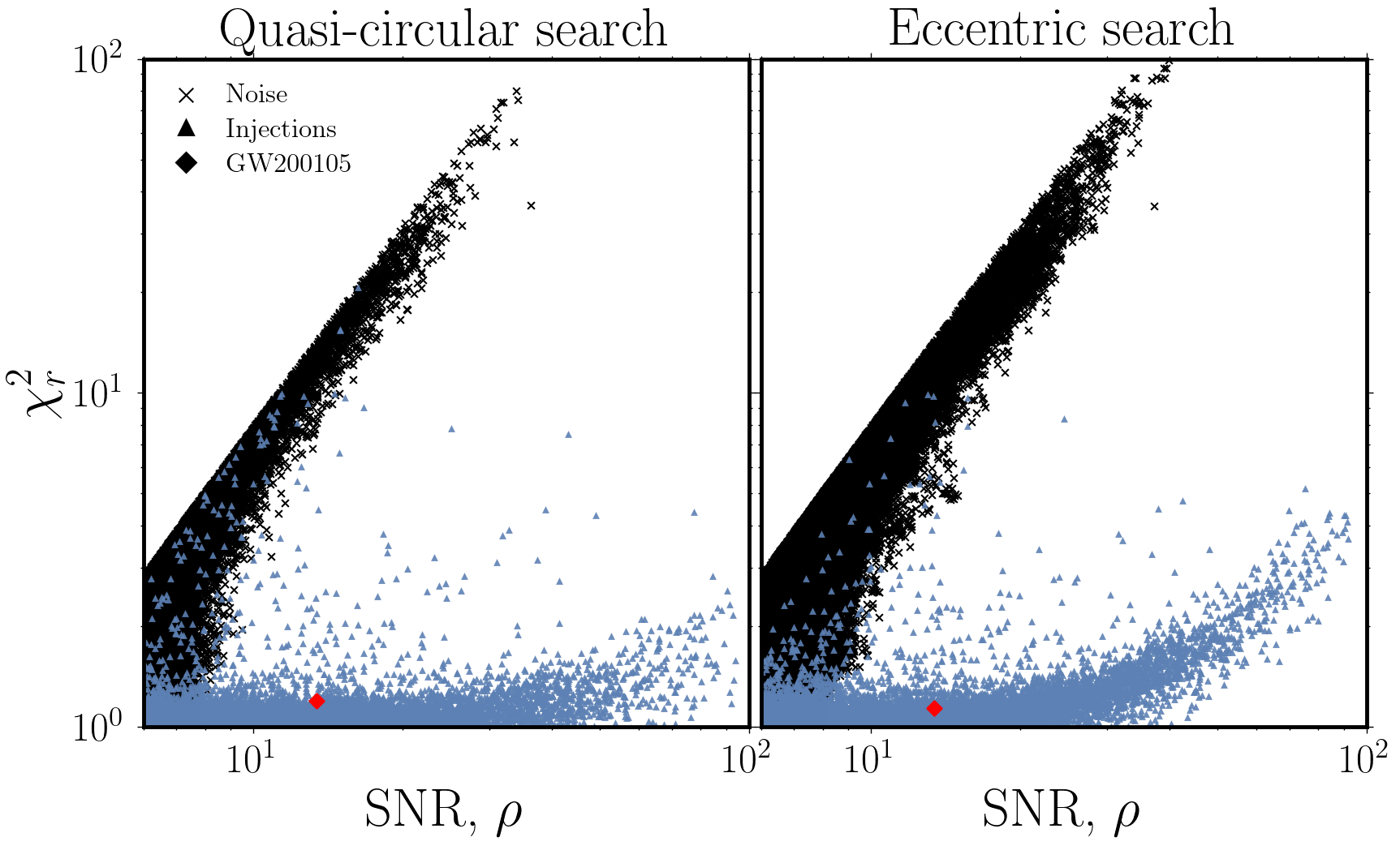} 
    \caption{\label{fig:snr_chisq_qc} The $(\rho, \chi_r^2)$-values of the quasi-circular injections (blue triangles) obtained in the quasi-circular (left) and eccentric (right) searches. Noise triggers (black cross marks) and the triggers corresponding to GW200105 (red diamonds) are the same as in Fig.~\ref{fig:snr_chisq}. In both Figs.~\ref{fig:snr_chisq} and \ref{fig:snr_chisq_qc} the noise triggers are well separated from the recovered injections and the GW200105 trigger in the respective searches.}
\end{figure*}

Both sets of injections are added to the data and analyzed with the eccentric and quasi-circular template banks. Searches compute various statistics, including the SNR $\rho$, the reduced chi-squared statistics $\chi_r^{2}$, and the IFAR values for recovered injections.   

In Fig.~\ref{fig:snr_chisq}, we present the SNR $\rho$ and reduced chi-squared $\chi_r^{2}$ values of the eccentric injections found in at least two detectors, and noise triggers of each detector from both searches. The noise triggers shown are found during the coincident operation of at least two detectors with $\rho > 6$ in each. Noise triggers found in the quasi-circular and eccentric searches are represented by the black cross symbols in each panel. Injections recovered by the search are shown by filled triangles pointing upward in each panel. These injections are colored by their eccentricity values at 20 Hz, $e_{20}^{}$. The $(\rho, \chi_r^{2})$-values for quasi-circular injections, along with those for noise triggers are shown separately in Fig.~\ref{fig:snr_chisq_qc}.

A search is considered robust if an injection within its parameter space is missed only when the signal is severely affected by data quality issues.
In our injection study, we do not find any instances of obviously detectable injections that are missed by either search. The populations of noise triggers are clearly distinguishable from the recovered quasi-circular and eccentric injections in the $\rho-\chi_r^{2}$-plane at sufficiently high $\rho$, indicating the robustness of the analysis. Almost all recovered injections have $\chi_r^{2}<10$, while noise triggers exhibit much higher values of $\chi_r^{2}$, resulting in a clear separation of noise triggers from recovered injection in the $\rho-\chi_r^{2}$ plane. 

The $\chi_r^{2}$ statistic for a signal in Gaussian noise attains its minimum value of 1 when the signal exactly matches a template. Larger $\chi_r^{2}$-values indicate either that the signal does not match the template or that the data is highly non-Gaussian. In general, a template does not exactly match a signal due to the finite spacing of the template bank. An additional mismatch occurs if the signal and the template differ in their physics content, for example if only one of them includes the effects of eccentricity.  
As shown in Fig.~\ref{fig:snr_chisq}, we find that the quasi-circular search recovers eccentric injections with higher $\chi_r^{2}$-values compared to the eccentric search. Conversely, the eccentric search yields higher $\chi_r^{2}$-values for quasi-circular injections than the quasi-circular search, as illustrated in Fig.~\ref{fig:snr_chisq_qc}. Since the simulated signals in both searches are injected into the same data and the two template banks are constructed with the same template spacing or minimal match of $0.97$, the observed trend of larger $\chi_r^{2}$-value when the waveform of an injected signal does not match that of the template, can only be attributed to a systematic difference between injected signals and templates.   
This systematic behavior is also reflected in the case of GW200105 in both searches. GW200105 is recovered with a similar SNR of $\sim 13.4$ in both searches, but with a best-matched template with a $\chi_r^{2}$ value of 1.20 in the quasi-circular search versus 1.14 in the eccentric search. The $(\rho, \chi_r^{2})$-values for GW200105 from the eccentric and quasi-circular searches are indicated by filled diamond markers in Figs.~\ref{fig:snr_chisq} and~\ref{fig:snr_chisq_qc}. As depicted in both figures, the event is clearly separated from the respective background triggers in each search.

As shown in Eq.~\eqref{eq:reweighted_snr}, at a given SNR, an increase in $\chi_r^{2}$ will result in a decrease in the reweighted SNR value $\hat{\rho}$, thereby down-weighting the ranking statistics $\Lambda_s^{}$. This effect is also observed for eccentric injections recovered by the quasi-circular template bank. The systematic down-weighting of eccentric injections in the quasi-circular search has a direct implication for the sensitivity of the search as it leads to a reduction of the number of eccentric detections above a given IFAR threshold in a quasi-circular search compared to an eccentric search. This is directly reflected in the $\langle VT \rangle$ of the searches. 

\begin{figure}[!t]
    \centering
    \includegraphics[width=0.48\textwidth]{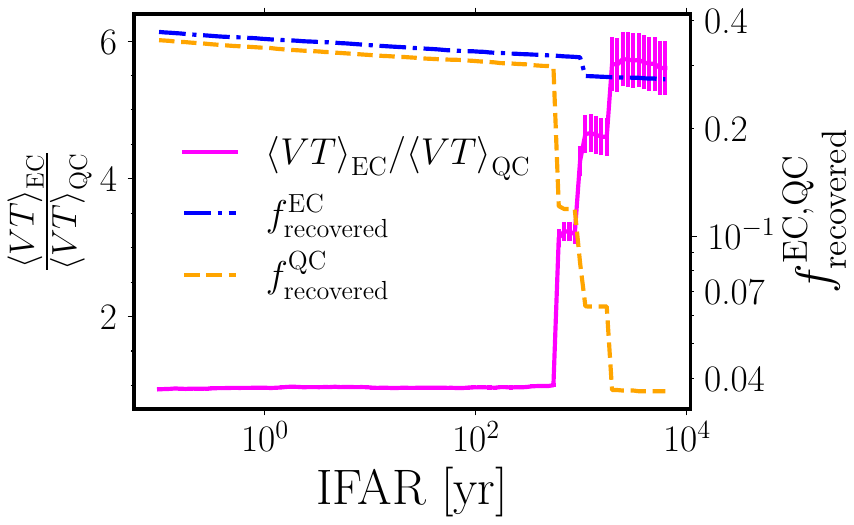}
    \caption{\label{fig:vt_ratio} The ratio of the sensitive spacetime volume between the eccentric bank $\langle VT \rangle^{}_{\rm EC}$ and the quasi-circular bank $\langle VT \rangle^{}_{\rm QC}$, estimated using eccentric injections at various IFAR thresholds on the left y-axis. The error bar of each point denotes the $1\sigma$-uncertainty. The fraction of eccentric injections recovered by the quasi-circular and eccentric searches are shown on the right y-axis, denoted by $f_{\rm recovered}^{\rm QC}$ and $f_{\rm recovered}^{\rm EC}$, respectively. The eccentric search is significantly more sensitive to eccentric signals than the quasi-circular search beyond an IFAR of $\sim$ 500 years. }
\end{figure}

To quantify the impact of neglecting eccentricity on the search sensitivity, we calculate the ratio of the $\langle VT \rangle$ between the two searches, $\langle VT \rangle^{}_{\rm EC}/\langle VT \rangle^{}_{\rm QC}$, by analyzing eccentric injections only, where the subscript
EC denotes the $\langle VT \rangle$ of the eccentric search and QC the one of the quasi-circular one. 
As mentioned earlier, each recovered injection shown in Fig.~\ref{fig:snr_chisq} is assigned a different IFAR value. We compute the ratio of sensitive spacetime volume using different IFAR thresholds for injection recovery. In addition, for each IFAR threshold we compute the fraction of eccentric injections recovered by the eccentric and quasi-circular searches, providing a complementary measure of the relative search performance. The recovery fractions of injections in the quasi-circular and the eccentric searches are denoted by $f_{\rm recovered}^{\rm QC}$ and $f_{\rm recovered}^{\rm EC}$, respectively. As shown in Fig.~\ref{fig:vt_ratio}, both searches perform comparably up to an IFAR of $\sim 500$ year. For higher IFAR thresholds, the eccentric search significantly outperforms the quasi-circular one with a $\langle VT \rangle$ that is 4 to 6 times larger than than that of the quasi-circular search. We find that the increase in sensitivity is mainly concentrated in regions of higher eccentricity with $e^{}_{20} \geq  0.1$, consistent with expectations based on results presented in~\cite{Phukon:2024amh}.

Additionally,  we compare the computational cost of the two searches. Matched filtering the data with the quasi-circular bank from the LIGO Hanford, LIGO Livingston, and the Virgo detectors required $1,834$, $1,683$, and $1,754$ CPU hours, respectively. For the eccentric bank, the corresponding CPU hours are $25,597$ for LIGO Hanford, $24,929$ for LIGO Livingston, and $26,637$ for Virgo. Across the detector network, the eccentric search is approximately 15 times more computationally expensive than the quasi-circular search, primarily due to the order of magnitude larger number of templates in the eccentric bank.

\subsection{Search results}
Both the quasi-circular and eccentric searches successfully recover GW200105 as a LIGO Livingston only event with high statistical significance. GW200105 is the most significant event in both searches, and other lower ranked events are consistent with noise fluctuations. Although the Virgo detector was in observing mode during the event, GW200105 was not observed in Virgo due to its low sensitivity at that time. The LIGO Hanford detector was not observing during the occurrence of the event. This makes the detection of GW200105 in our analyses consistent with its classification as a single-detector event in GWTC-3~\cite{KAGRA:2021vkt,LIGOScientific:2021qlt}.

\begin{table*}[!ht]
\caption{\label{tab:search_summary} Summary of results for GW200105 from both searches: Value of the ranking statistics $\Lambda^{}_s$, exclusive IFAR in years, best-matching template parameters $\left(m_1^{}, m_2^{}, s_{1z}^{}, s_{2z}^{}, e_{20}^{}\right)$, SNR $\rho$, reduced chi-squared value $\chi_r^{2}$ and reweighted SNR $\hat{\rho}$ from quasi-circular and eccentric  searches. }
\begin{ruledtabular}
\begin{tabular}{c c c c c c c c c c c c}
search & $\Lambda_s^{}$ & IFAR [yr]  & $m_1^{}$ $[M^{}_{\odot}]$ & $m_2^{}$ $[M^{}_{\odot}]$ & $s^{}_{1z}$ &  $s^{}_{2z}$ & $e_{20}^{}$ &  $\rho$ &$\chi_r^2$ & $\hat{\rho}$ \\
\hline\\
quasi-circular  & 26.43 & 976  & 8.14 & 2.29 & -0.16 & 0.05 & N/A & 13.42 & 1.20 &12.47 \\
\hline \\
eccentric & 27.47 & $\geq$1000 &  9.18 & 2.06 & -0.08	& 0.05	 & 0.11 & 13.41 & 1.14 & 12.64  \\
\end{tabular}
\end{ruledtabular}
\label{tab:candidate}
\end{table*}

In our targeted quasi-circular search, GW200105 is detected with an SNR $\rho$ of 13.42, a ranking statistic $\Lambda_s^{}$ of 26.43 and an exclusive IFAR of $976$ years. This IFAR value corresponds to a false alarm probability (FAP) of $2.28 \times 10^{-5}$, where the FAP is the probability of finding one or more noise background events exceeding the  ranking statistics value of GW200105 in the observation time~\cite{Usman:2015kfa}. In the eccentric search,  GW200105 is detected with a comparable SNR $\rho$ of $13.41$, but up-ranked to a higher ranking statistics  $\Lambda_s^{}$ of 27.47. These translate to a lower bound of the IFAR of 1000 years  and a lower FAP of $2.22 \times 10^{-5}$ in the eccentric search. Following~\cite{LIGOScientific:2016vbw}, FAPs are converted to single-sided Gaussian standard deviations $\sigma$ according to 
\be
\sigma =-\sqrt{2}\, {\rm erf}^{-1}\{ 1-2(1-{\rm FAP}) \},
\ee
where ${\rm erf}^{-1}$ is the inverse error function. The significance of GW200105 in the eccentric and quasi-circular searches is $4.083 \sigma$ and $4.077 \sigma$, respectively. When the 1000-year IFAR-cap in the extrapolation of the FAR for a single detector event is removed in both searches, the eccentric search assigns GW200105 an IFAR of  27417 years, while the quasi-circular IFAR remains unchanged at 976 years.

\begin{figure*}[!t]
    \centering
    \includegraphics[width=0.7\textwidth]{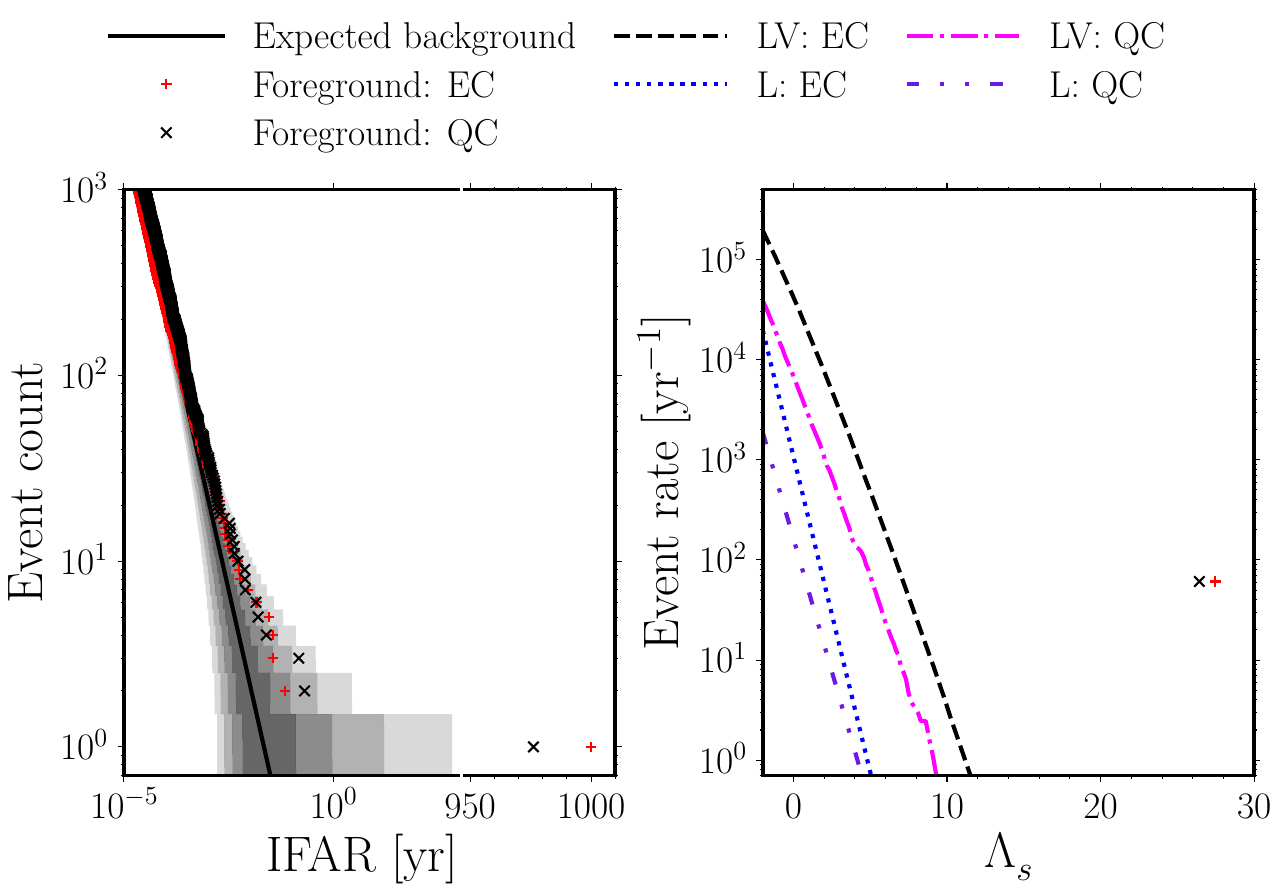}
    \caption{\label{fig:search-results} 
    {\it Left}: Points with black cross and red plus marks denote the number of  triggers  with exclusive IFAR greater than or equal to a given value in foregrounds of the quasi-circular and eccentric searches, respectively. The solid black line is the number of  expected background events, which is the same for both searches as the live-times are the same for both searches. The gray bands indicate the $1\sigma$ to $4\sigma$ uncertainties. The $x$-axis is a mix of log and linear scales to showcase differences in the significance of GW200105 between the two searches. 
    {\it Right}: The rate of events ranked louder or equal to the ranking statistics $\Lambda_s^{}$ in the LV networks' exclusive background for the two searches given by graphs labeled as LV:EC and LV:QC, respectively. The curves with labels L:EC and L:QC denote the same quantity for the LIGO Livingston only background for the eccentric and the quasi-circular searches. The black cross and red plus symbol give the rate of events at the ranking statistics of GW200105 in the foreground of respective searches.}
\end{figure*}

We emphasize that the significances of GW200105, as reported above in terms of FAP- and $\sigma$-values should be interpreted cautiously. The conversion of the IFAR to FAP and subsequently to $\sigma$-values depends explicitly on the analyzed livetime of the detectors or the total foreground time~\cite{Usman:2015kfa}. Since our targeted analysis is conducted over a limited data segment with a total foreground time of approximately $8.11$ days surrounding GW200105, the reported FAP- and $\sigma$-values should be considered an indicative measure derived from the IFAR values of GW200105. For these reasons, we present IFAR values as the primary metric for assessing the statistical significance of GW200105. A summary of our searches is provided in Tab.~\ref{tab:search_summary} and Fig.~\ref{fig:search-results}.

It may seem surprising that GW200105 is more significant in the eccentric search than in the quasi-circular one, particularly given that the background is comparatively louder in the eccentric search than the quasi-circular one as shown in the right panel of Fig.~\ref{fig:search-results}. However, the improvement in significance can be understood using the results of the injection campaign carried out in Sec~\ref{sec:vt}. As shown in Fig.~\ref{fig:snr_chisq}, although both template banks can recover eccentric injections, eccentric injections filtered against the quasi-circular bank tend to produce systematically larger $\chi^2_r$ values, leading to a stronger down-weighting of $\rho$ to produce lower values of reweighted SNR $\hat{\rho}$.  
This reduces the ranking statistic in the quasi-circular bank  search compared to eccentric bank search, thereby lowering significance.  We see consistent behavior for triggers with properties similar to GW200105 in both searches. The $\chi^2_r$-value for GW200105 in the quasi-circular search is slightly larger than in the eccentric search. The $\left(\rho, \chi^2_r \right)$-values obtained for GW200105 in the quasi-circular and eccentric searches are shown by the filled diamond in left and right panels of Figs.~\ref{fig:snr_chisq} and~\ref{fig:snr_chisq_qc}. 

\begin{figure}
    \centering
    \includegraphics[width=0.48\textwidth]{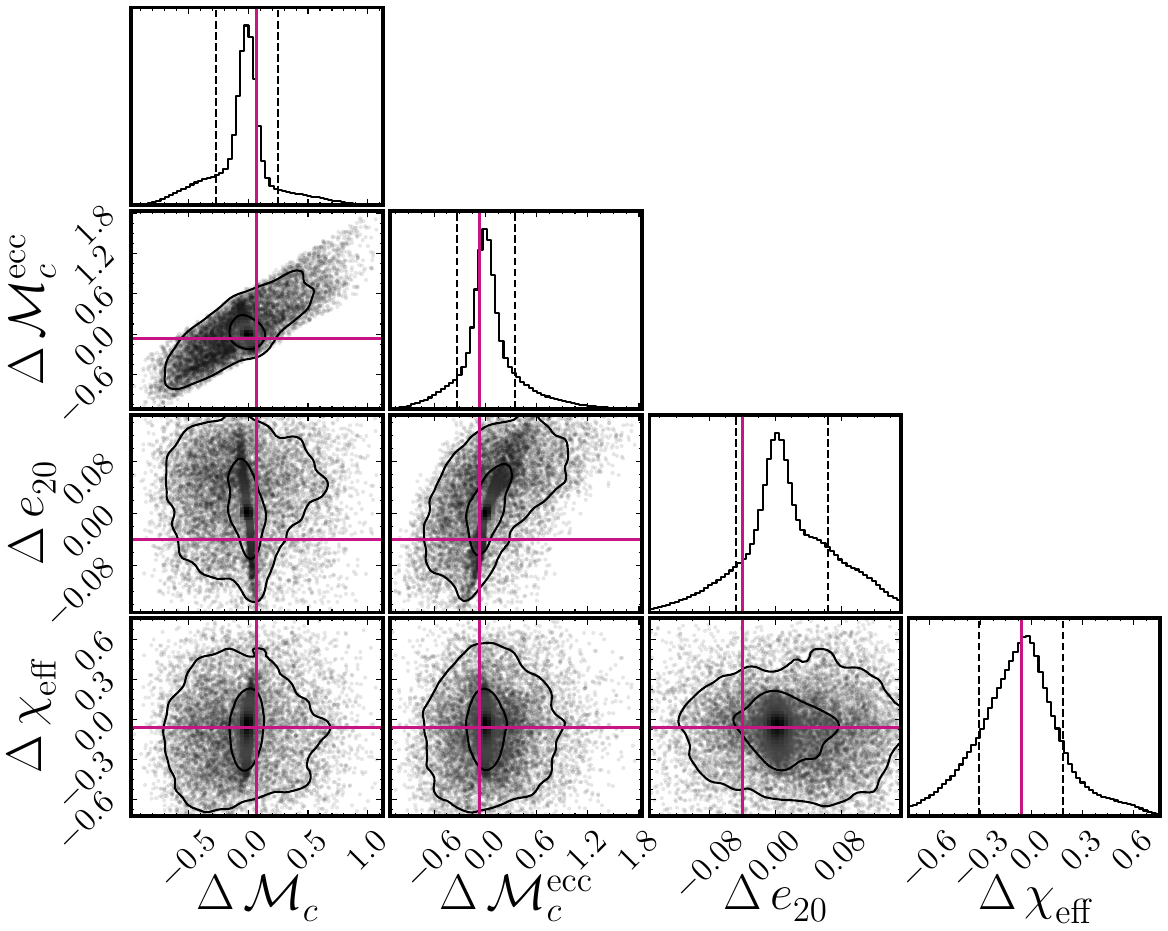}
    \caption{\label{fig:param_offset} Differences $(\Delta)$ between parameters of eccentric injections and their best-matching templates in the eccentric search. Contours denote the 50\% and 90\% confidence intervals, while dashed vertical lines in the one-dimensional marginalized distributions indicate $1\sigma$ regions. Solid lines denote the offset for GW200105 in the eccentric search relative to parameter estimation results of~\cite{Morras:2025xfu}.}
\end{figure}

To assess whether the detection of GW200105 in the eccentric search by the template given in Table~\ref{tab:search_summary} reflects a genuine response of the template bank rather than a random fluctuation, we perform two investigations. 
For each eccentric injection recovered by the eccentric search described in Sec.~\ref{sec:vt}, we compute the differences between the injection parameters and those of the corresponding best-matching template. The resulting distributions of these parameter offsets benchmark the expected template-signal mismatch in the search. We then compute the parameter offsets for GW200105 relative to the template that recovered the event, using the median posterior values from~\cite{Morras:2025xfu} as the true parameters of the source. As shown in Fig.~\ref{fig:param_offset}, the offsets for GW200105 lie within the $1\sigma$ region of the injection recovery offset distribution. Therefore, the separation between the template and true parameters for GW200105 is statistically indistinguishable from the offsets produced when the search successfully recovers genuinely eccentric injections.

\begin{figure}
    \centering
    \includegraphics[width=0.49\textwidth]{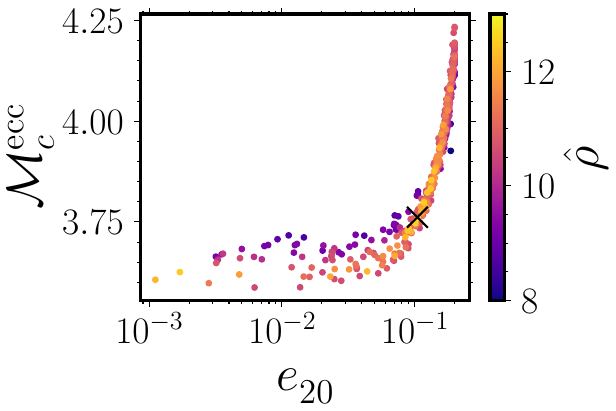}
    \caption{\label{fig:GW200105_triggers} Reweighted SNRs $\hat{\rho}$ of triggers in the LIGO Livingston data from the eccentric search within $\pm 1s$ around GW200105, shown as a function of $e_{20}$ and $\mathcal{M}_c^{ecc}$. Triggers with $\chi_r^{2} > 1.5$ and SNRs $\rho < 8$ are not included in the figure. The black cross marks the trigger assigned to GW200105, which has the largest $\hat{\rho}$ value in the cluster.  
     }
\end{figure}

Next, we examine the relationship between the reweighted SNR, $\hat{\rho}$, and the template eccentricity $e_{20}^{}$ in the cluster of triggers obtained from the LIGO Livingston data within $\pm 1s$ around the trigger time of GW200105. Any signal generally produces a cluster of triggers because it matches different templates to a varying degree. For an astrophysical signal, the triggers in this cluster should follow a meaningful pattern reflecting the underlying signal properties. In case of an eccentric signal, we expect $\hat{\rho}$ to increase with template eccentricity towards the eccentricity of the trigger that is assigned against the signal in the search. Such a trend is highly atypical in noise induced triggers or triggers caused by quasi-circular signals. In the cluster of triggers around GW200105, $\hat{\rho}$ is preferentially higher in triggers with significant eccentricity than for those with near-zero eccentricity. In Fig.~\ref{fig:GW200105_triggers} we show triggers from the cluster around GW200105 with SNR $\rho > 8$ and $\chi_r^{2}<1.5$ in the $\mathcal{M}_c^{\rm ecc} - e_{20}^{}$-plane, with the colors indicating the values of $\hat{\rho}$. The observed behavior in parameter offset of GW200105 and nearby triggers make GW200105 qualitatively consistent with an eccentric signal.

\section{Discussion}
\label{sec:discussion}

In this paper, we presented a targeted eccentric inspiral matched-filter search for the NSBH event GW200105 using the PyCBC pipeline. The evidence of eccentricity in GW200105 as identified in the Bayesian parameter estimation analysis of~\cite{Morras:2025xfu}, provides an opportunity to test if GW searches can be made more sensitive to binaries with residual eccentricity in the detectors' band by including more physical effects in the search space. Our analysis leveraged recent developments in search methods, such as the construction of an eccentric template bank construction for low mass binaries~\cite{Phukon:2024amh} and the estimation of significance for single-detector events in PyCBC~\cite{Davies:2022thw}. A quasi-circular search is also performed over the same non-eccentric parameter space to provide a direct comparison of performance between the two searches.

In recent years, several eccentric matched-filter analyses have been performed on LIGO-Virgo data, each targeting distinct regions of parameter space, but none have reported statistically significant detections~\cite{Nitz:2019spj,Nitz:2021mzz,Dhurkunde:2023qoe,Pal:2023dyg,Wang:2025yac}. These analyses did either not consider single-detector triggers to identify candidate events, or GW200105 lies outside their search parameter space, particularly in the eccentric NSBH search of~\cite{Dhurkunde:2023qoe}.
In our targeted eccentric search, GW200105 is detected as a high-significance single-detector event in the LIGO-Livingston data. While the quasi-circular search also detects the event, the comparison between the two analyses shows that GW200105 attains a higher significance in the eccentric search, reaching the maximum allowed IFAR value for single detector events, corresponding to a significance greater than $4.083 \sigma$. GW200105 is also detected in the quasi-circular search, with a significance of $4.077 \sigma$. 

The comparable significance for GW200105 obtained by the two searches is a non-trivial result. The FAR of a templated search is expected, at leading order, to scale approximately with the number of templates $N^{}_{\rm templates}$ reflecting the increased trials factor, FAR $\propto N^{}_{\rm templates}$~\cite{Keppel:2013yia}. The eccentric bank used in this work is approximately nine times larger than the quasi-circular bank, resulting in a much larger search background than for the quasi-circular one. The better  significance of GW200105 in the eccentric search compared to the corresponding value from the quasi-circular search is attributed to better signal consistency between GW200105 and the templates. Furthermore, when the 1000-year cap imposed on the IFAR of a single detector event is removed, the eccentric search assigns GW200105 an IFAR exceeding the capped value by more than an order of magnitude, while the IFAR from the quasi-circulare search remain unchanged. If the quasi-circular bank was made denser to match the eccentric bank's size and the densified bank's search retains the same ranking and detection statistics values as the original quasi-circular search, then the corresponding IFAR of GW200105 would be reduced approximately by a factor of nine to $\sim 109$ yr. The reported increase in significance of GW200105 with the eccentric search is particularly encouraging from a binary population perspective, as nearly $10\%$ of NSBH binaries formed in hierarchical triples can retain residual eccentricity $e_{20}^{}$ as large as $0.1$ at a GW frequency of 20 Hz~\cite{Stegmann:2025clo}. 

Besides the search results, a key output of a search analysis is the sensitivity of the search to a population of compact binaries, which is provided in terms of sensitive spacetime volume $\langle VT \rangle$~\cite{Essick:2025zed}. 
Analyses on simulated eccentric signals in both searches showed that the quasi-circular search has a significant selection bias against binaries with eccentricities $e^{}_{20} \geq 0.1$. Such binaries are less much effectively recovered by a quasi-circular search. Incorporating eccentricity into the search significantly reduces this selection bias, resulting in up to a sixfold improvement in the sensitive $\langle VT \rangle$ compared to the quasi-circular search. An accurate quantification of the selection function for compact binary observations, accounting for all physical effects including eccentricity and spin-induced precession, is essential for conducting hierarchical Bayesian analyses to characterize the properties of the source population. The quantification of the selection function for eccentric binaries is a less explored avenue~\cite{Zevin:2021rtf}, and we leave accurate modeling of the selection function for eccentric sources for future work.

Finally, our study highlights the importance of extending compact binary searches beyond the quasi-circular assumption to encompass eccentric signals. The search results motivate the continued development and integration of eccentric search pipelines into GW analyses, particularly in the context of future observing runs where increasingly sensitive detectors provide the opportunity to uncover a broader range of astrophysical sources.

\section*{Acknowledgments}
The authors thank Thomas Dent for helpful comments and suggestions. K.S.P., G.P. and P.S. acknowledge support from STFC grant ST/V005677/1; G.P. and P.S. also acknowledge support from STFC grant ST/Y00423X/1.
G.P. is very grateful for support from a Royal Society University Research Fellowship URF{\textbackslash}R1{\textbackslash}221500 and RF{\textbackslash}ERE{\textbackslash}221015, and UK Space Agency grant ST/Y004922/1. G.P. gratefully acknowledges support from an NVIDIA Academic Hardware Grant. P.S. also acknowledges support from a Royal Society Research Grant RG{\textbackslash}R1{\textbackslash}241327. P.S. and G.P. are grateful to the Institute for Advanced Study for hospitality, where parts of this work were carried out. The authors are grateful for computational resources provided by the LIGO Laboratory (CIT) and supported by the National Science Foundation Grants PHY-0757058 and PHY-0823459, the University of Birmingham's BlueBEAR HPC service, which provides a High Performance Computing service to the University's research community, the Bondi HPC cluster at the Birmingham Institute for Gravitational Wave Astronomy, as well as resources provided by Supercomputing Wales, funded by STFC grants ST/I006285/1 and ST/V001167/1 supporting the UK Involvement in the Operation of Advanced LIGO. This research has made use of data or software obtained from the Gravitational Wave Open Science Center (gwosc.org), a service of the LIGO Scientific Collaboration, the Virgo Collaboration, and KAGRA. This material is based upon work supported by NSF's LIGO Laboratory which is a major facility fully funded by the National Science Foundation, as well as the Science and Technology Facilities Council (STFC) of the United Kingdom, the Max-Planck-Society (MPS), and the State of Niedersachsen/Germany for support of the construction of Advanced LIGO and construction and operation of the GEO600 detector. Additional support for Advanced LIGO was provided by the Australian Research Council. Virgo is funded, through the European Gravitational Observatory (EGO), by the French Centre National de Recherche Scientifique (CNRS), the Italian Istituto Nazionale di Fisica Nucleare (INFN) and the Dutch Nikhef, with contributions by institutions from Belgium, Germany, Greece, Hungary, Ireland, Japan, Monaco, Poland, Portugal, Spain. KAGRA is supported by Ministry of Education, Culture, Sports, Science and Technology (MEXT), Japan Society for the Promotion of Science (JSPS) in Japan; National Research Foundation (NRF) and Ministry of Science and ICT (MSIT) in Korea; Academia Sinica (AS) and National Science and Technology Council (NSTC) in Taiwan. This manuscript has the LIGO document number P2500673.

\bibliography{ref}

\end{document}